\documentclass[a4paper,final]{iopart}

\usepackage{amssymb}
\usepackage{dcolumn}
\usepackage{graphicx}
\usepackage{longtable}
\usepackage{bm}
\usepackage{color}
\usepackage{cite}

\usepackage{graphicx}% Include figure files
\usepackage{iopams}
\usepackage{epsfig}

\begin{document}

\title{Half metallic ferromagnetism in tri-layered perovskites Sr$_4$T$_{3}$O$_{10}$ (T=Co, Rh)} 

\author{Madhav P. Ghimire$^{1,2}$, R. K. Thapa$^{3}$, D. P. Rai$^{3}$, Sandeep$^{3}$,  T. P. Sinha$^{4}$ and Xiao Hu$^{2}$}
\address{$^{1}$ Nepal Academy of Science and Technology, P. O. Box 3323, Khumaltar, Lalitpur, Nepal}
\address{$^{2}$ International Center for Materials Nanoarchitectonics (WPI-MANA), National Institute for Materials Science, Tsukuba 305-0044, Japan}
\address{$^{3}$ Department of Physics, Mizoram University, Aizawl 796-004, India}
\address{$^{4}$ Department of Physics, Bose Institute, Kolkata 796-004, India}
\ead{Hu.Xiao@nims.go.jp}

\begin{abstract}
     First-principles density functional theory (DFT) is used to investigate the electronic and magnetic properties of Sr$_4$Rh$_3$O$_{10}$, a member of the Ruddlesden-Popper series. 
Based on the DFT calculations taking into account the co-operative effect of Coulomb interaction ($U$) and spin-orbit couplings (SOC), Sr$_4$Rh$_3$O$_{10}$ is found to be a half metallic ferromagnet (HMF) with total angular moment $\mu_{\rm {tot}}$=12$\mu_B$ per unit cell. The material has almost 100$\%$ spin-polarization at the Fermi level despite of sizable SOC.
Replacement of Rh atom by the isovalent Co atom is considered. Upon full-replacement of Co, a low-spin to intermediate spin transition happens resulting in a HMF state with the total angular moment three-time larger (i.e. $\mu_{\rm {tot}}$=36$\mu_B$ per unit cell), compared to Sr$_4$Rh$_3$O$_{10}$. 
We propose Sr$_4$Rh$_3$O$_{10}$ and Sr$_4$Co$_3$O$_{10}$ as  candidates of half metals.
\end{abstract}


\maketitle

\section{Introduction}
The emerging developments of magneto-electronic devices have intensified  research on novel materials that can be applied as a single-spin electron sources and high-efficiency magnetic sensors\cite{wolf}. One of the key points to achieve such materials is to manipulate the spin polarization by band engineering. 
Half metals (HMs) are one class of promising materials which are metallic in one spin channel, while insulating in the opposite spin channel due to the asymmetric band structures. HMs can generate fully spin-polarized currents without any external operation\cite{groot1,pickett1}.
 Ideally HM is specified by integer spin magnetization in units of Bohr magneton, where the total number of electrons per unit cell is an integer and all valence bands are fully filled in the insulating spin channel. HM states have been identified in several groups of materials which includes oxides (CrO$_2$), mixed-valent manganites (La$_{1-x}$A$_x$MnO$_3$ where A=Ca, Ba, Sr), Heusler alloys (NiMnSb), perovskites (Sr$_2$FeTO$_6$, where T=Mo, Re), etc.\cite{schwarz,dowben,pickett2,richter,kobayashi,sarma,mpg,
mpg1,felser,galanakis,kandpal,muller,katsnelson,xiao1,xiao2}.

A recent study on layered-perovskites belonging to the Ruddlesden-Popper (RP) series\cite{popper} exhibits novel properties desirable in spintronics\cite{felser}. 
In cobaltates, the single-layered compound Sr$_2$CoO$_4$ shows ferromagnetic (FM) as well as HM behavior\cite{matsuno,pandey}. The same material is reported to show negative magnetoresistance\cite{wang}. La$_{1+x}$Sr$_{2-x}$Mn$_2$O$_7$ is found to exhibit colossal magnetoresistance (CMR)\cite{moritomo} while its nonstoichiometric samples  shows novel microstructural features\cite{seshadri}. 
In ruthenates, the change in number $n$ of RuO$_2$ layers %in Sr$_{n+1}$Ru$_n$O$_{3n+1}$
 leads to a variety of collective phenomena: spin-triplet chiral superconductivity \cite{maeno} ($n=1$); heavy $d$-electron masses \cite{lee} ($n=2,3$), CMR effects \cite{lin} ($n=2$) and itinerant ferromagnetism and metamagnetism ($n=3$)\cite{cao,mao} as well as HM ferromagnetism ($n=\infty$)\cite{jeng}.
Furthermore, the delicate interplay between electron-electron correlation and spin-orbit coupling (SOC) in  iridates gives rise to various unconventional phases. 
Insulator-metal transition was observed in Sr$_{n+1}$Ir$_n$O$_{3n+1}$ with increasing $n$\cite{moon}. Sr$_2$IrO$_4$ and Sr$_3$Ir$_2$O$_7$ have been realized as a $J_{\rm {eff}}=1/2$ Mott-insulator evidenced by an experiment\cite{moon,kim1,kim2,okada,zhang} whereas, SrIrO$_3$ is a paramagnetic metal\cite{zhang}. 

A newly synthesized tri-layered rhodate Sr$_4$Rh$_3$O$_{10}$ \cite{yamaura} belonging to Sr-Rh-O family in the RP series\cite{popper} comes into our attention with the unique properties.  The crystal with general formula Sr$_{n+1}$Rh$_n$O$_{3n+1}$  (where $n=3$) is built up by regular intergrowth of rock-salt Sr-O sheets and perovskites (SrRhO$_3$)$_n$. Rh atoms carry on charge state $+4$ with spin magnetizations, and the material exhibits strong crystal distortion due to rotation and tilting of the octahedra. With a total of 68 atoms in its unit cell, the system has four formula units. 
The other member in the series includes well studied compounds such as Sr$_2$RhO$_4$, Sr$_3$Rh$_2$O$_7$, and SrRhO$_3$\cite{itoh,lee2,haverkort,kim3,yamaura1,yamaura2}. 

In this study, we investigate for the first-time the electronic and magnetic properties of Sr$_4$Rh$_3$O$_{10}$ by using first-principles density functional theory (DFT).
The ground-state of Sr$_4$Rh$_3$O$_{10}$ is found to be a half-metallic ferromagnet (HMF). With an integer total angular moment $\mu_{\rm {tot}}=12\mu_B$ per unit cell, the HM state is robust to strong electron-correlation and SOC effects. 
Replacement of Rh by Co atoms is considered in Sr$_4$Rh$_3$O$_{10}$. For full replacement, the material Sr$_4$Co$_3$O$_{10}$ is a HMF with total magnetization increasing by three times of magnitude due to intermediate spin configuration. The insulating state of Sr$_4$Co$_3$O$_{10}$ flips to the opposite spin channel  in contrast with Sr$_4$Rh$_3$O$_{10}$.

\section{Crystal structures and methods}
 Sr$_4$Rh$_3$O$_{10}$ crystallizes in an orthorhombic structure with space group $Pbam$ as shown in figure~1. The unit cell is composed of triple layers of corner-shared RhO$_6$ octahedra separated by double rock-salt layers of Sr-O. Each layer is formed independently by two types of Rh atoms as shown in figure~1. In the crystal of Sr$_4$Rh$_3$O$_{10}$ two in-equivalent oxygen atoms were found along the plane (O1) and along the apex (O2). The bond-lengths between Rh and planar (1.98$\rm {\AA}$) and apical (2.02$\rm {\AA}$) oxygens are different due to tilting and rotation of RhO$_6$ octahedron. The amplitude of rotation of RhO$_6$ octahedra estimated from experiments\cite{yamaura} is $\sim$12$^\circ$ which is close to that of Sr$_3$Rh$_2$O$_{7}$ (10.5$^\circ$)\cite{yamaura1}.
 \begin{figure}[t]
    \centering
    \psfig{figure=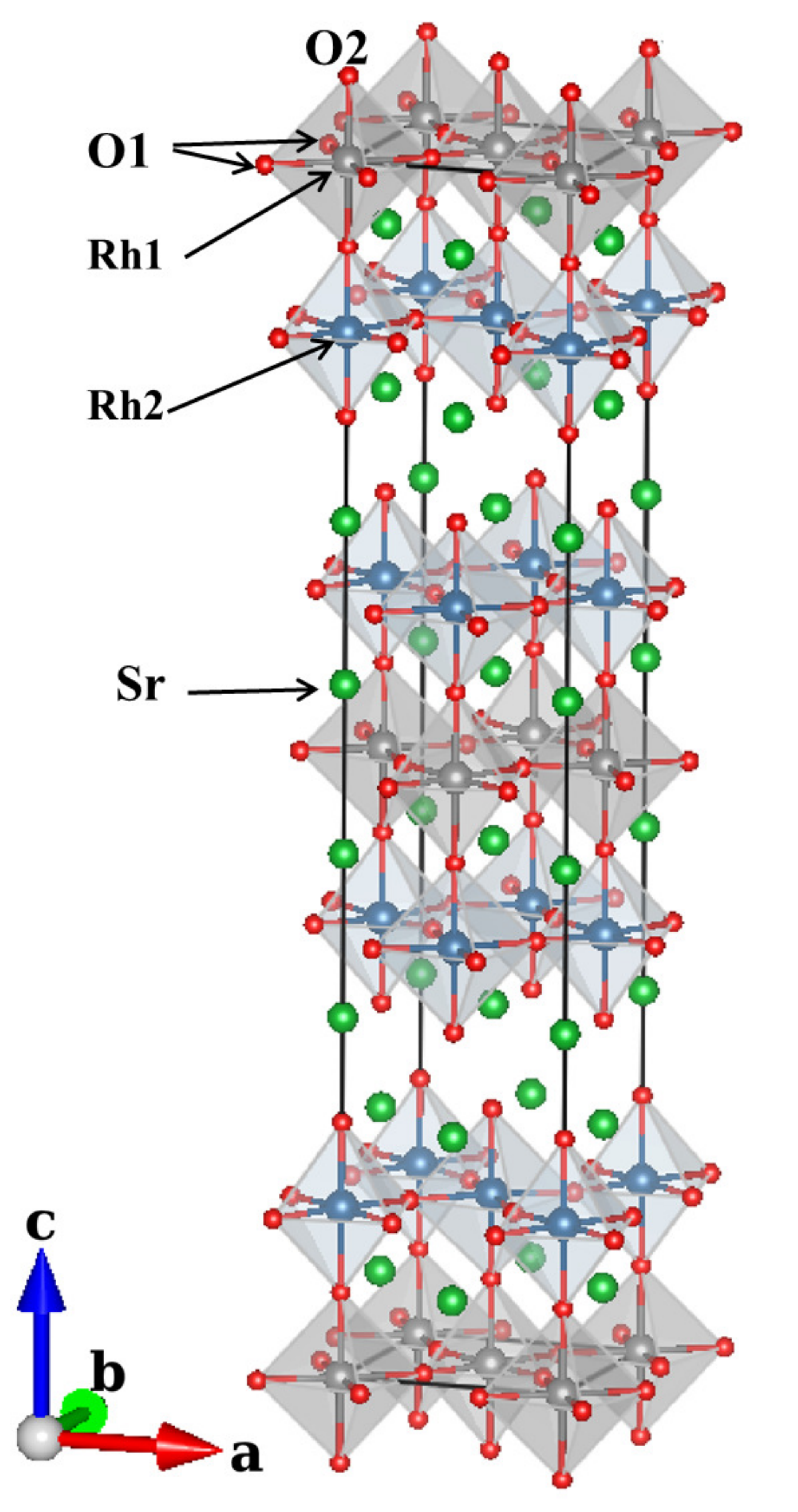,width=1.7in,height=3.4in}
     \caption{Crystal structure of the tri-layered perovskite Sr$_4$Rh$_3$O$_{10}$. O1 (planar) and O2 (apical) are the oxygen bonded to Rh ions forming RhO$_6$ octahedron.} 
     \label{Fig. 1}
  \end{figure}

The electronic structure calculations were performed by using the full-potential linearized augmented plane wave plus local orbital method implemented in the WIEN2k code \cite{blaha}. The atomic sphere radii ({\it{R}}$_{\rm {MT}}$) were 2.2, 2.07, 1.97 and 1.7 Bohr for Sr, Rh, Co, and O respectively. The standard generalized gradient approximation (GGA) exchange-correlation potential within the Perdew-Burke-Ernzerhof scheme
 \cite{perdew} were used. 
The results presented below are for Coulomb interaction\cite{lichtenstein} $U$=3 and 5eV for Rh and Co, respectively, and we have confirmed that the main conclusions do not change with $U=1\sim5$eV for Rh and $U=4.2\sim8$eV for Co\cite{note}.
The SOC was considered via second variational method using the scalar-relativistic orbitals as basis.
Starting from the experimental results \cite{yamaura}, we optimize the lattice parameters to determine the equilibrium positions of all the individual atoms and reach the stable structure using WIEN2k code \cite{blaha} with the force convergence set at 1mRy/a.u. All the presented results are of the optimized structure tabulated in table~1, which agrees with the results obtained  from the experimental lattice parameters\cite{yamaura}. 
\begin{table}[t]
  \centering
  \caption{Optimized atomic coordinates for Sr$_4$Rh$_3$O$_{10}$ obtained by minimizing the total energy from first principles DFT calculations.}
  
  \begin{tabular}{r c c c}
  \br
{ Atom} &    $x$    &  $y$    &  $z$   \\\cline{1-4}
   Sr1  & 0.5   & 0.0  & 0.0697   \\
   Sr2  & 0.5   & 0.0   & 0.2047 \\
   Sr3  & 0.0   & 0.0   & 0.2953 \\
   Sr4  & 0.0   & 0.0   & 0.4303 \\
   Rh1  & 0.0   & 0.0   & 0.0 \\
   Rh2  & 0.0   & 0.0   & 0.14 \\
   Rh3  & 0.5   & 0.0   & 0.354 \\
   Rh4  & 0.5   & 0.0   & 0.5 \\
   O1  & 0.3135   & 0.1866   & 0.0 \\
   O2  & 0.1919   & 0.3081   & 0.1391 \\
   O3  & 0.3080   & 0.3081   & 0.3609 \\
   O4  & 0.1865   & 0.1866   & 0.5 \\
   O5  & 0.0   & 0.0   & 0.0689 \\
   O6  & 0.0   & 0.0   & 0.21 \\
   O7  & 0.5   & 0.0   & 0.2899 \\
   O8  & 0.5   & 0.0   & 0.4311 \\
 \br  
     \end{tabular}

  \label{tab:1}
 \end{table}

\section{Parent material 
  S\lowercase{r}$\mathbf{_4}$R\lowercase{h}$\mathbf{_3}$O$\mathbf{_{10}}$}

In the ionic picture, Rh with five $d$-electrons nominally has a charge state $+4$. The five 4$d$ electrons occupy the $t_{2g}$ triplet and leave the higher $e_g$ doublet empty due to large crystal-field generated by oxygen octahedron. The low-spin alignment of Rh is then $t_{2g}^3\uparrow$, $t_{2g}^2\downarrow$ giving rise to a moment of 1$\mu_B/$atom.

In order to know the ground state electronic and magnetic structures of Sr$_4$Rh$_3$O$_{10}$, we perform the DFT calculations considering the co-operative effect of $U$ and SOC. The ground state is found to be ferromagnetic (FM) with the lowest energy among other magnetic configurations.
To compare the relative effect from $U$ and SOC, the total density of states (DOS) are plotted in figure~2 with GGA, GGA$+U$ and GGA+$U$+SO functional.
Within GGA functional, Sr$_4$Rh$_3$O$_{10}$ shows metallic characters for both  spin-up and spin-down channels.
For $U\geq1$eV, the topmost valence band in spin-up channel shifts deeper towards the valence region giving rise to an insulating state. 
 In contrary, for spin-down channel the metallic state is maintained, except that the DOS peak near $E_{\rm F}$ is splitted due to strong electron-correlations. The effect of $U$ is comparatively small as expected due to partial occupation of the orbitals in spin-down channel. With spin-up channel insulating (band-gap ~0.9eV) and the spin-down channel metallic, Sr$_4$Rh$_3$O$_{10}$ is a half-metal (HM).
It is generally considered that the presence of heavy element with strong SOC strength might destroy the HM state. In the present case it turns out that the HM state is fully preserved even with SOC as observed in figure~2(c). This happens due to the large-exchange splitting between the spin-up and spin-down bands at $E_{\rm F}$. 
\begin{figure}[t]
     \centering
    \psfig{figure=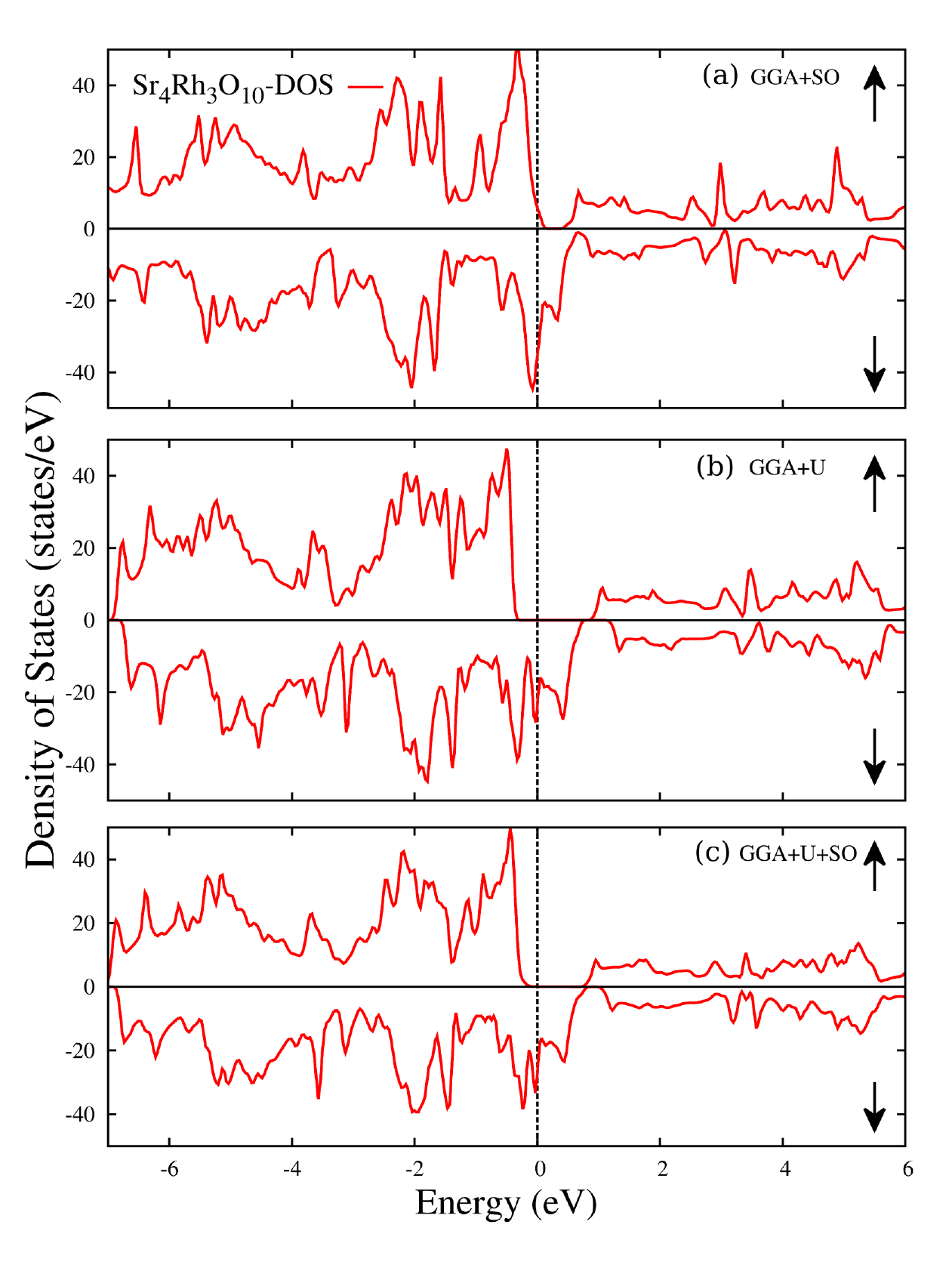,width=3.2in,height=4.2in}
     \caption{Total DOS for Sr$_4$Rh$_3$O$_{10}$ within (a) GGA+SO, (b) GGA+$U$, and (c) GGA+$U$+SO, respectively.}
    \label{pDOS}
\end{figure}

\begin{figure}[t]
    \centering
    \psfig{figure=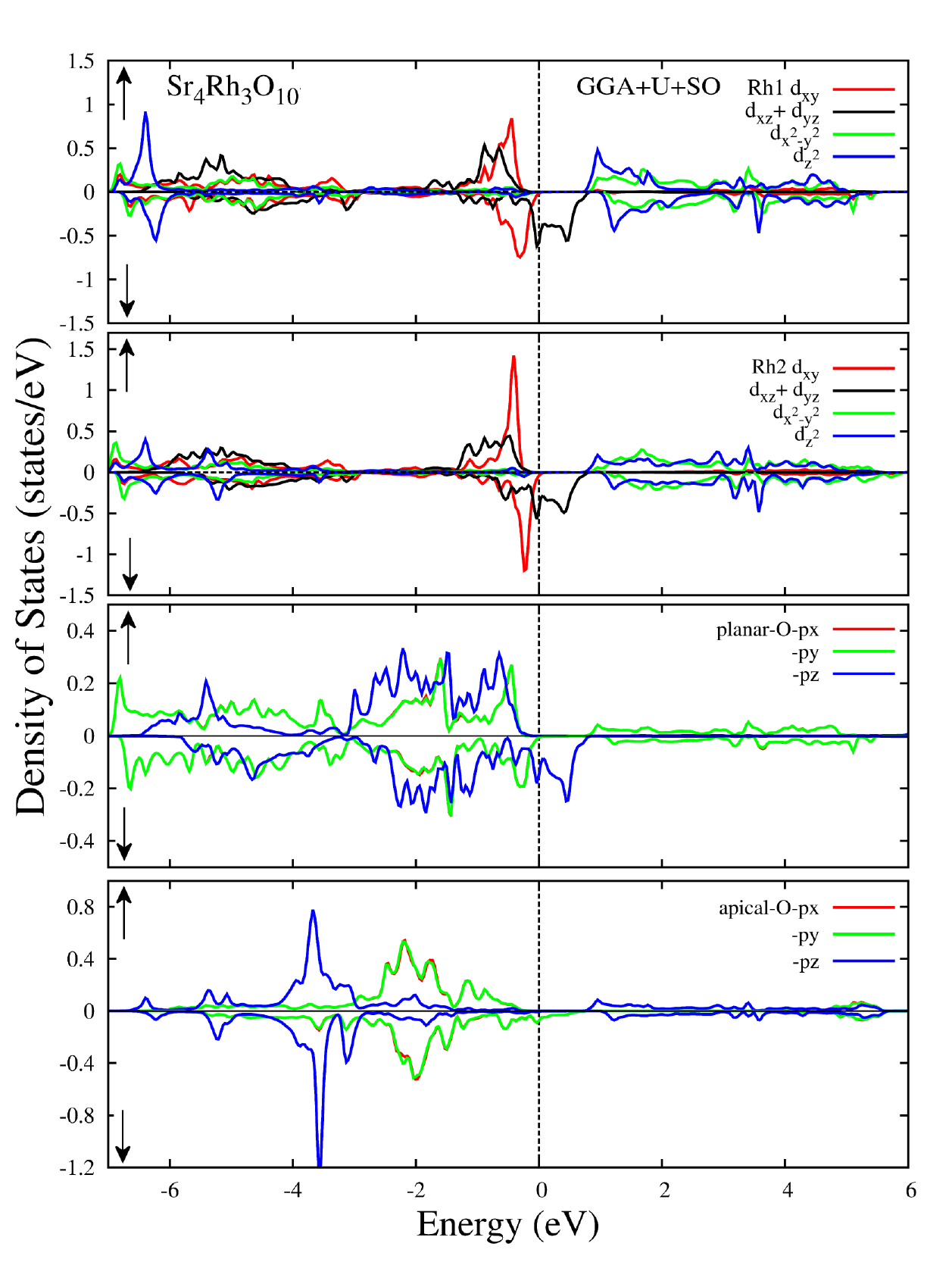,width=3.2in,height=4.2in}
     \caption{Partial density of states for two in-equivalent  Rh-4$d$ states and contributions from planar as well as apical oxygen-2$p$ states in spin-up ($\uparrow$) and spin-down ($\downarrow$) channels for Sr$_4$Rh$_3$O$_{10}$ within GGA+$U$+SO.}
    \label{Fig. 3}
\end{figure}

As revealed by the site-projected DOS shown in figure~3 for spin-up and spin-down channels obtained from first-principles calculations within GGA+$U$+SO, two in-equivalent Rh ions play significant role in dictating the electronic properties of Sr$_4$Rh$_3$O$_{10}$. 
The crystal field effect is found prominent in Rh which causes the splitting between $t_{2g}$ and $e_g$ states.
The partial DOS contribution from $d$ states of two in-equivalent Rh atoms shows full occupation of the $t_{2g}$ states with its sub-level $d_{xy}$ and $d_{xz}+d_{yz}$ lying below $E_{\rm F}$, whereas the $e_g$ states are in the conduction region for spin-up channel. For spin-down channel, the $t_{2g}$ states are partially occupied due to deficiency of one electron in the $d_{xz}+d_{yz}$ sub-levels. As a result, these bands cross $E_{\rm F}$ and form a continuous band at $E_{\rm F}$. The order of occupation observed in figure~3 is consistent with the ionic picture. The crystal-field energy ($\sim$1.3eV) is close to the exchange energy splitting ($\sim$1.2eV) between spin-up and spin-down bands of Rh atoms. This clearly indicates the low-spin configuration consistent with the ionic picture. 

As shown in figure~3, Rh1 has a dominant $t_{2g}$ characters near $E_{\rm {F}}$ with an itinerant DOS due to its strong hybridization with the oxygen atoms. On the other hand Rh2 has a sharp peak near $E_{\rm F}$ contributed by $d_{xy}$ states since the hybridization with the oxygen is relatively weak unlike the $d_{xz}+d_{yz}$ states. This occurs mainly due to crystal distortion in Sr$_4$Rh$_3$O$_{10}$. Since the hybridization of Rh with the planar oxygen (O1) is stronger than apical oxygen (O2)  in both spin channels, the moment induced at O1 is larger than O2 (see table~2). 

\begin{figure}[t]
    \centering
    \psfig{figure=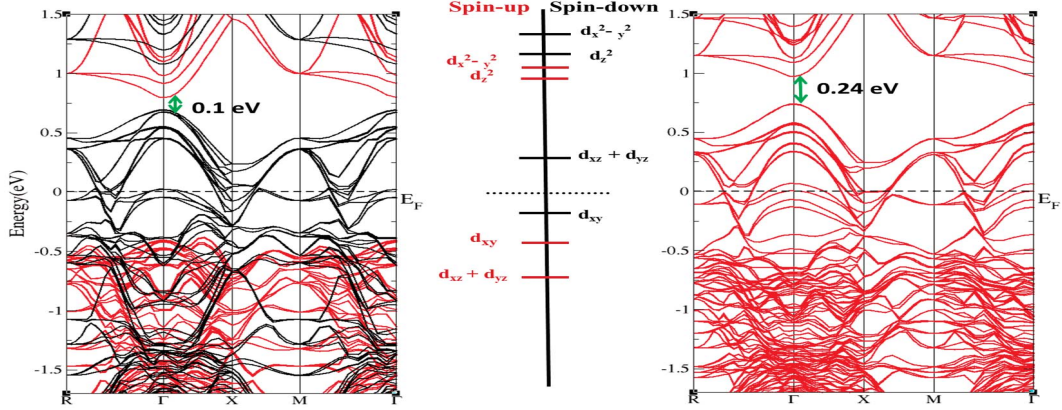,width=5.5in,height=2.7in}
    \caption{Band structures for Sr$_4$Rh$_3$O$_{10}$ within  GGA+$U$ (left) and GGA+$U$+SO (right) approach. Schematic band-diagram (middle) demonstrates the contribution from Rh-$d$ bands in the band structure for GGA+$U$. Bands in red (black) within GGA+$U$ indicates spin up (down) channel.}
    \label{Fig4}
\end{figure}

To distinguish the effect of SOC, we show the spin-resolved band structure for Sr$_4$Rh$_3$O$_{10}$ within GGA+$U$ and GGA+$U$+SO as shown in figure~4, together with a schematic diagram of the band ordering of Rh-$d$ states. First we look at the gap between spin-up and spin-down channel in the conduction region for GGA+$U$. The gap is found to be $\sim$0.1eV at high symmetry point $\Gamma$ in the Brillouin zone (BZ) (see figure~4(left)). When SOC is considered, the electronic band structure is modified. The major changes are observed in the conduction region above $0.7$eV and in the valence region below -0.3eV. Due to SOC effect of  Rh ($\zeta=$191meV)\cite{haverkort}, the spin-up bands in the conduction region at $\sim0.7$eV are pushed up towards higher energy by the spin-down bands. This generates a gap of $\sim0.24$eV at the high symmetry point $\Gamma$  in the conduction region (see figure~4(right)).  Similar changes are observed below  $E_{\rm {F}}$ around $-0.3$eV, where the spin-up bands are pushed deeper in the valence region. Thus, the bands at $E_{\rm F}$ are purely of spin-down channel. This shows that HM state is fully preserved in Sr$_4$Rh$_3$O$_{10}$ even though sizable SOC exists.

At the ground state obtained from first-principles calculations, the total angular moment ($\mu_{\rm tot}$) is 12$\mu_{\rm B}$ per unit cell contributed by Rh atoms and oxygen atoms (see table~2).
The orbital contribution in Rh ($\sim0.1\mu_{\rm B}$) is in accordance with the Hund's third rule, where Rh with more than half-filled $t_{2g}$-shell has its orbital moment parallel to its spin moment.
The Rh moment is less than unity, with the remaining part transferred to oxygen due to strong hybridization as seen from the partial DOS (see figure~3). Planar oxygen atoms have higher moment than the apical oxygen due to shorter bond-length with Rh-ions. 
In an ionic picture, each Rh ion carries moment $+1\mu_{\rm B}$ giving rise to $\mu_{\rm tot}$ $=12\times(+1\mu_{\rm B})=12\mu_{\rm B}$ in a unit cell, consistent with the first-principles calculations.

\begin{figure}[t]
    \centering
       \psfig{figure=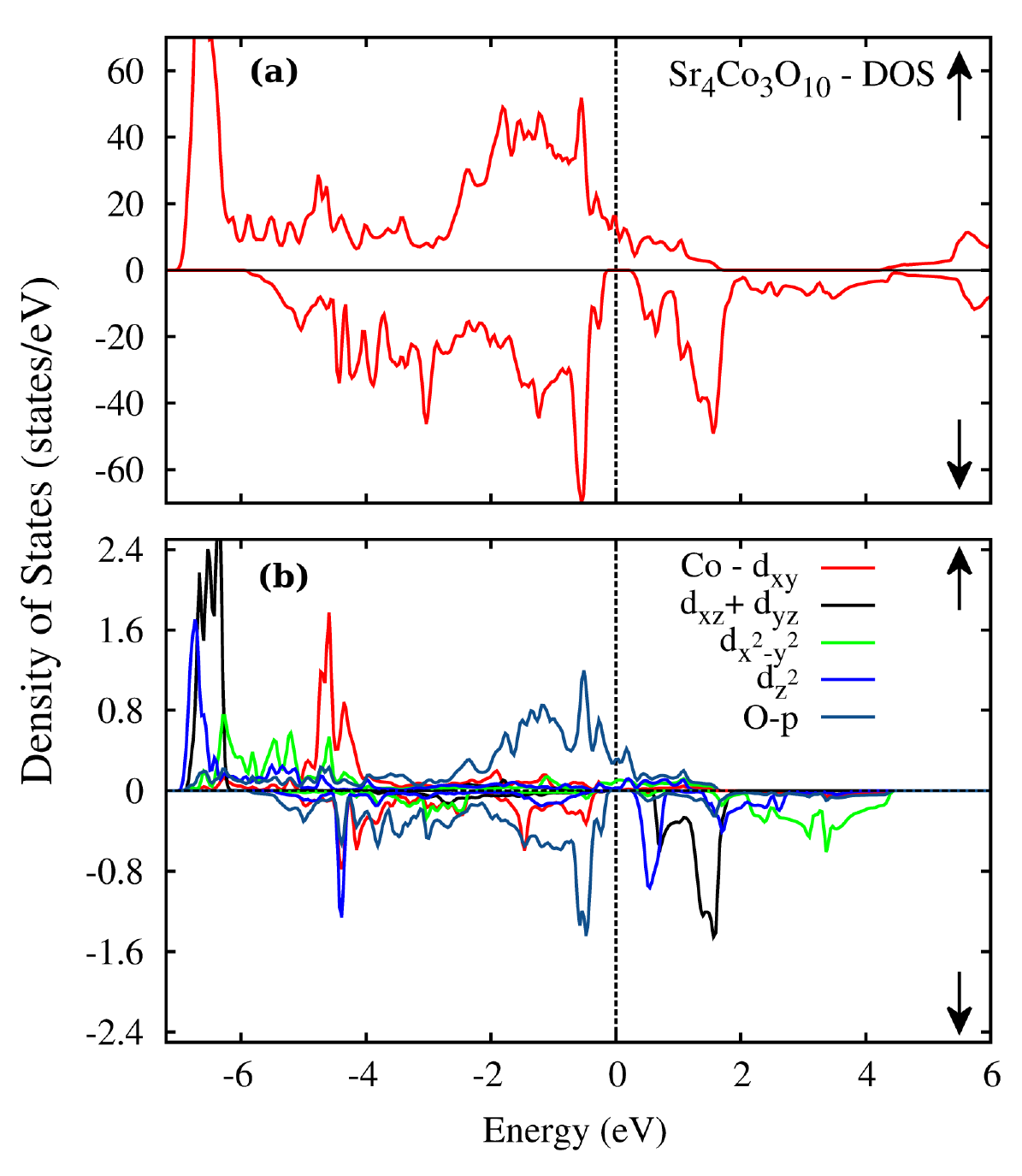,width=3.2in,height=4.in}
 	    \caption{Density of states for Sr$_4$Co$_3$O$_{10}$ in spin-up ($\uparrow$) and spin-down ($\downarrow$) channels within GGA+$U$+SO: (a)  total; (b)  partial contribution from Co-$d$ and O-$p$orbitals.}
    \label{Fig. 5}
\end{figure}

\begin{table}[t]
  \centering
  \caption{Moments per atom of Rh and Co, one set of two in-equivalent oxygen atoms and unit cell ($\mu_{\rm tot}$)
   for Sr$_4$T$_3$O$_{10}$ (where T=Rh, Co) from first-principles calculations within GGA+$U$+SO. The unit of moments is the Bohr magneton $\mu_{\rm B}$. The contributions from individual atoms are within muffin-tins while the total moment includes those from the interstitial regime.}

  \begin{tabular}{r c c c c c}
  \br
{ } &    T1    &  T2    &  O1   &   O2  &  {$\mu_{\rm tot}$} \\\cline{1-6}
   Sr$_4$Rh$_3$O$_{10}$  & 0.63   & 0.54  & 0.11   &  0.05  & 12 \\
   Sr$_4$Co$_3$O$_{10}$  & 2.65   & 2.68   & 0.02   & 0.18   & 36 \\
   \br
  \end{tabular}

  \label{tab:1}
 \end{table}

\section{Doped material 
  S\lowercase{r}$\mathbf{_4}$C\lowercase{o}$\mathbf{_3}$O$\mathbf{_{10}}$}

We consider the replacement of Rh by Co atoms, which corresponds to iso-valent doping into the parent material Sr$_4$Rh$_3$O$_{10}$.   In the ionic picture Co retains the charge state $+4$. The spin state transforms from low-spin of Sr$_4$Rh$_3$O$_{10}$ to intermediate-spin $t_{2g}^3\uparrow,{d^1_{3z^2-r^2}}\uparrow,d_{xy}^1\downarrow$. The spin-state transition is observed frequently in cobaltates\cite{eschrig}.
 
We focus on the most interesting case (i.e. full-replacement of Rh by Co atoms) where HM state is achieved.  

From the total and partial DOS of Sr$_4$Co$_3$O$_{10}$ shown in figure~5,  
 the material has full occupation of the $t_{2g}$ states and partial occupation of the $e_g$ states resulting to a metallic state in spin-up channel. In spin-down channel, $d_{xy}$ state  is fully occupied whereas the remaining $t_{2g}$ and $e_g$ states lie in the conduction region. This gives rise to band-gap of $\sim$0.5eV in spin-down channel. 
 Due to crystal distortion the $d$ states of Co hybridize strongly with the O-2p states in both spin-channels. 
 It is interesting to note that doping of Co flips the insulating state from spin-up to spin-down channel due to intermediate-spin configuration of Co in Sr$_4$Co$_3$O$_{10}$\cite{note}. 
Thus, with  metallic state in spin-up channel and the insulating state in spin-down channel, the material Sr$_4$Co$_3$O$_{10}$ is a HM. 

From the first-principles calculations, the total angular moment is found to be $\mu_{\rm tot}$ = $36\mu_{\rm B}$ per unit cell. 
The moments of Co atoms and oxygen atoms are shown in table~2.

 \section{Discussions}
First-principles calculations on magnetic anisotropy energy of Sr$_4$Rh$_3$O$_{10}$ and Sr$_4$Co$_3$O$_{10}$ indicate the $c$ axis of the crystal as their easy axis with anisotropy energy of $\sim3$meV  and $\sim0.95$meV per unit cell, respectively.
%-600.8403303Ry along 001
%-600.8402618Ry along 100

   SOC mixes the spin-up and spin-down electrons together which possibly destroys the HM state. In the present materials the exchange-splitting between the spin-up and spin-down DOS at $E_{\rm F}$ is large (see figures~2 and ~5), which makes the mixing marginally small at $E_{\rm F}$. Hence, HM state is preserved even with SOC.

From our calculations, the material Sr$_4$Rh$_3$O$_{10}$ has larger exchange splitting which reduces the mixing of spin-up and spin-down bands at $E_{\rm F}$, and its low total moment causes a small stray field generated by local magnetic ions. Hence compared to Sr$_4$Co$_3$O$_{10}$ with a three-time larger moment, Sr$_4$Rh$_3$O$_{10}$ is more desirable as a HMF.

\section{Conclusions}
Based on the first-principles density functional approach, Sr$_4$Rh$_3$O$_{10}$ is found to be a half metallic ferromagnet  
due to the cooperative effect from Coulomb interaction, spin-orbit coupling and the crystal field. It is demonstrated that doping iso-valent atom into the system by replacing Rh with Co, a half metal with larger moment is achieved. It is emphasized that the large exchange splitting between spin-up and spin-down bands at the Fermi level retains the half metallicity even in presence of spin-orbit coupling.
%\section*{Acknowledgments}
\ack{
    MPG thanks M. Richter, IFW-Dresden for valuable discussions and suggestions and the organizers of FPLO-2013 for financial support to present this work. MPG acknowledges the partial support from NAST (Nepal) and INSA (India) under the NAST-INSA bilateral exchange program. This work was supported partially by WPI Initiative on Materials Nanoarchitectonics, MEXT, Japan.}


\begin{thebibliography}{99}
\bibitem{note}
   It should be noted that band gap opens for Sr$_4$Co$_3$O$_{10}$ in spin-down channel for $U_{\rm eff}\geq4.2$eV (where $U_{\rm eff}=U-J$ with $J$ fixed to 0.8eV).
\end{thebibliography}
\end{document}